\documentclass[10pt,conference]{IEEEtran}
\renewcommand\IEEEkeywordsname{Keywords}
\newcommand{\squeezeup}{\vspace{-2.5mm}}
\ifCLASSINFOpdf
\else
\fi
\usepackage{graphics}
\usepackage{graphicx}
\usepackage{listings}
\usepackage{verbatim}
\usepackage[frozencache,cachedir=.]{minted}
\usepackage{xparse}
\usepackage{amsmath}
\usepackage{dirtytalk}
\usepackage{amssymb}
\usepackage{caption}
\begin{document}
\title{PAARS: Privacy Aware Access Regulation System}

\author{\IEEEauthorblockN{Md. Monowar Anjum}
\IEEEauthorblockA{Department of Computer Science\\University of Manitoba\\
anjumm1@myumanitoba.ca}
\and
\IEEEauthorblockN{Noman Mohammed}
\IEEEauthorblockA{Department of Computer Science\\University of Manitoba\\
noman@cs.umanitoba.ca}
}

\IEEEoverridecommandlockouts
\IEEEpubid{\makebox[\columnwidth]{Copyright: 978-1-5386-5541-2/18/\$31.00~\copyright2018 IEEE \hfill} \hspace{\columnsep}\makebox[\columnwidth]{ }}

\maketitle
\IEEEpubidadjcol

\begin{abstract}
\boldmath
During pandemics, health officials usually recommend access monitoring and regulation protocols/systems in places that are major activity centers. As organizations adhere to those recommendations, they often fail to implement proper privacy requirements to prevent privacy loss of the users of those protocols or systems. This is a very timely issue as health authorities across the world are increasingly putting these regulations in place to mitigate the spread of the current pandemic. A number of solutions have been proposed to mitigate these privacy issues existing in current models of contact tracing or access regulations systems. However, a prevalent pattern among these solutions are they mainly focus on protecting users privacy from server side and involve Bluetooth based ephemeral identifier exchange between users. Another pattern is all the current solutions try to solve the problem in city-wide or nation-wide level. In this paper, we propose a system, PAARS, which approaches the privacy issues in access monitoring/regulation systems from a micro level. We solve the privacy issues in access monitoring/regulation systems without any exchange of any ephemeral identifiers between users. Moreover, our proposed system provides privacy on both server side and the user side by using secure hashing and differential privacy mechanism.   
\end{abstract}



% \fi
%%\IEEEpeerreviewmaketitle

\section{Introduction}
Pandemics often spread by contact events and therefore, require contact tracing by government or public health officials to mitigate the spread of the pandemic. Public health officials also recommend access regulation/restriction for institutions with higher activity level such as banks, educational institutions, religious organizations etc. While these organizations adhere to the regulations set by the public health authorities strictly, they often fail to understand the value of privacy when they take these recommended steps.

In pre-pandemic era, the topic of privacy was extensively explored in literature. However, privacy issues in contact tracing systems didn't attract much attention from the academia. With the unprecedented spread of the pandemic, governments and public health officials of a number of countries resorted to digital contact tracing to mitigate the spread of the pandemic which resulted in a new debate over privacy requirements and citizen's rights both in mainstream media and academia. Arundhati Roy, 1997 Man Booker prize winner for \textit{“God of Small Things”} famously quoted recently in the wake of the government contact tracing apps: “\textit{Pre-corona, if we were sleepwalking into the surveillance state, now we are panic-running into a super-surveillance state.}” It may sound like an alarmist plea. However, prevalence of access monitoring at organizational level and contact tracing applications at national level do support her statement as most of the time these systems do not maintain proper privacy requirements needed.

For instance, In order to comply with the building occupancy limit set by the health officials, the University of Manitoba in Winnipeg, Canada recently implemented an access monitoring system. This step involves everyone who intend to access the campus buildings register their intended time and reason of visit prior to visiting. Another example is the MIA's (Manitoba Islamic Association) recent guideline to limit the capacity of all mosques during each prayer time and implement an online registration system in order to ensure that. Let's take a look at more concrete examples, to understand the privacy concerns presented by systems like the ones mentioned in the preceding statements. 

\begin{table}[t]
    \caption{Sample Raw Access Log Data of Two Users}
    %\centering
    \begin{tabular}{|c|c|c|c|c|}
    \hline
    User & Entry & Exit & Location & Reason \\
    \hline
    X & Aug 13,2020 15:30 & Aug 13,2020 18:30& EITC-2 & Lab-1 work \\
    \hline
    Y & Aug 13,2020 15:00 &Aug 13,2020 19:30 & EITC-2 & Lab-1 work \\
    \hline
    X & Aug 16,2020 10:30 &Aug 16,2020 18:30 & EITC-2 & Lab-1 work \\
    \hline
    Y & Aug 16,2020 15:00 &Aug 16,2020 20:30 & EITC-2 & Lab-1 work \\
    \hline
    X & Aug 17,2020 15:30 &Aug 17,2020 16:30 & EITC-2 & Lab-1 work \\
    \hline
    Y & Aug 17,2020 10:00 &Aug 17,2020 19:30 & EITC-2 & Lab-1 work \\
    \hline
    Y & Aug 19,2020 12:00 &Aug 19,2020 19:30 & EITC-2 & Lab-1 work \\
    \hline
    \end{tabular}
    \label{tab:sample_monitoring_data_uofm}
\end{table}

\begin{table}[t]
    \caption{Sample Raw Religious Institution Attendance Data}
    \centering
    \begin{tabular}{|c|c|c|c|c|c|}
    \hline
    User & Prayer 1 & Prayer 2 & Prayer 3 & Prayer 4 & Prayer 5 \\
    \hline
    X & \checkmark & \checkmark & \checkmark & & \checkmark\\
    \hline
    Y & \checkmark & \checkmark & \checkmark & \checkmark & \checkmark\\
    \hline
    Z &  &  &  &  & \checkmark\\
    \hline
    \end{tabular}
    \label{tab:sample_monitoring_data_mia}
\end{table}
\textit{Example 1:} Table \ref{tab:sample_monitoring_data_uofm} represents an access monitoring system data for two users X and Y. It can be seen from the table that there were multiple encounters between two users over a given period of time and based on that it can be inferred with reasonable confidence that these two individual know and interact with each other. We consider this a privacy breach and in the subsequent sections of this work we call this as \textbf{exposure of social interaction graph} of an user.

\textit{Example 2:} Table \ref{tab:sample_monitoring_data_mia} represents raw access log data from a religious institution for three users X,Y and Z. Based on the frequency of their prayer attendance, a reasonable inference can be made about their level of religiousness which is a gross privacy violation.

Recent works on privacy preserving contact tracing applications in both industry and academia focused on the privacy of contact tracing/access regulation problem in a macro level. In most of these works, the users were to exchange some kind of pseudo-random tokens that can not be traced back to them via Bluetooth of their smartphone and the tokens would be later used to determine whether the user in question have been in contact with someone who were positively diagnosed for the disease. While these systems try to offer a trade off between strong privacy and utility, they are difficult to implement due to the infrastructure requirement and reluctance of users to adopt new technology. With no end of pandemic in sight, governments across the world are trying to reopen the economies around the world and allow the population to return to normal state of life. In order to do this, the spread of the pandemic must be controlled and digital contact tracing/access regulation with proper privacy can not be discarded as an option for that.

\textbf{Privacy needs:}\say{\textit{[Some of my patients] were more afraid of being blamed than dying of the virus}} - Lee Su-Young, Psychiatrist at Myongji Hospital, South Korea \cite{raskar2020apps}. This quote shows that the social stigma can be worse than the disease and the cost of privacy loss is much higher than generally perceived. Therefore, it is important to understand the privacy requirements of a system before designing it. Firstly, in the design of our proposed system, we intended to ensure that the system will not reveal any information to one user that helps him/her to determine the infection status of another user. Secondly, the system design and data flow must not reveal any sensitive data to the server so that the social interaction graph of any user can be deduced. Thirdly, the system design should not allow any malicious adversary to gain access to the communication between entities within the system. A detailed analysis on the security of the system design can be found in section \ref{sec:sec_analysis}.

\textbf{Information Needs:} Our proposed system design should be able to satisfy two core information needs. Government health officials have set the limit of building occupancy level during the pandemic to no more than 50\%. In order to comply with that regulation, the first information need from the system is the current state of the building occupancy level. The proposed system should be able to monitor this and alert authority if any intervention is required. The second information need is to enable the users to be alerted in case they have significant probability of being positively diagnosed for the disease. Once the users are notified about their probability of being positively diagnosed, they can decide to whether self-isolate or seek help from medical professionals. This action helps to minimize panic among all users as only users with significant probability will act on the information. In addition to these information needs, there are certain desirable traits in system design. For instance, the proposed system design should be practical. It should not require any addition to the infrastructure within the system environment which is discussed more in subsection \ref{sec:infrastructure}.  

\textbf{Contributions:} We decided to address the privacy issues in the access regulation and contact tracing systems at a micro level. We design a privacy-aware access monitoring/regulation system which can be implemented at organizational level without substantial change in the existing infrastructure while satisfying the privacy needs described above. Our key contributions are listed as follows:
\begin{itemize}
    \item We designed PAARS, an efficient access regulation system that provides strong privacy guarantee while considering existence of semi-honest server in our system model. The semi-honest entity in our model does not have access to any sensitive information that will allow it to deduce social interaction graph of any user or link any of the information available to itself to any particular user without the user's explicit consent.
    \item Our system design for PAARS requires minimal infrastructure addition as our system works on existing infrastructure available within the organization.
    \item We designed a novel probability calculation scheme to determine the positive diagnosis probability of the user. Our scheme uses the state of the art epidemiological models available in literature to calculate the disease exposure score of an user which is then used to calculate the probability of the user being positively diagnosed for the disease. The whole process for calculation of probability and query result release to the user is differentially private which preserves each user's privacy from any other users or attackers (Section \ref{sec:probability}). 
\end{itemize}
The rest of the paper is organized as follows. In the next section, we provide a background on contact tracing and differential privacy as well as provide with an overview of the recent works in these two domains. In the next section we present PAARS methodology and its working procedures. The sections that follow are utility analysis and security analysis which contains the privacy loss calculation and security vulnerabilities analysis for the PAARS system. We conclude the paper by giving a short analytic discussion on our approach.

\section{Background and Related Works}
\textbf{Contact/Proximity Tracing} is an umbrella term for the generic approach of identifying a contact event between two users of the system. Usually the systems use Bluetooth or GPS to perform the detection of contact between two users. When two users come in contact, they exchange a token which may or may not contain any persistent identifier of the user. These exchanged tokens can later be used to identify if there has been a contact with an infected user and take measures accordingly.

Recent works on contact tracing have focused heavily on privacy issues of the underlying principles of contact tracing. Epione \cite{trieu2020epione} is a lightweight contact tracing system proposed by Trieu et. al. which relies on faster private set intersection cardinality method to achieve efficiency over other similar methods. Their work specifically focuses on the case of matching between large scale contact database and small input queries. DP-3T \cite{dp20203t} is one of the most prominent one among the recent works on proximity tracing. This proposed system uses a decentralized model and provides robust privacy guarantees. PEPP-3T \cite{pepp20203t} is another similar protocol with centralized design instead of decentralized. Centralized systems offer more useful data which can be used by health authorities to make effective decisions however, numerous scholars believe that this approach might be harmful as it can become a mass surveillance tool for governments. EPIC \cite{altuwaiyan2018epic} is another similar scheme which provides a proximity tracing scheme by using hybrid wireless and localization technology. However, this system has scalability issues as noted in \cite{dar2020applicability}. 

There are multiple real world implementations of the contact tracing protocols. South Korean Government implemented their own contact tracing system which is reported to have widespread privacy issues \cite{korea2020ct}. Tracetogether \cite{bay2020bluetrace} is an adoption of the contact tracing protocol by the government of Singapore. It is a centralized system which stores the user phone number, identifying information and a randomized token. It does not store gps locations of the user. However, being a centralized system it is vulnerable to multiple security and privacy risks. Moreover, the slow adoption rate among the population also hinders it's applicability and effectiveness to a certain extent. In addition, Bluetooth based systems suffer from security vulnerabilities and privacy issues. Sophisticated GPS based solutions exist as noted in \cite{berke2020assessing} by Berke et. al. However, the computational overhead of this system prevents implementation by hindering scalability due to the fact that MPC (multi party computation) is used to preserve privacy of this system. 

Private Kit \cite{raskar2020apps} is another protocol proposed by Raskar et. al. which is a privacy-first contact tracing protocol by design. It provides a mix of voluntary sharing and unicasting which eliminates the need of a central monitoring entity. PACT \cite{chan2020pact} is another example of similar protocol with improvements over the earlier versions of contact tracing protocols available at the beginning of the pandemic. Technology industry leaders Apple and Google proposed their joint collaborative contact tracing protocol for smartphone devices which ensures strong privacy \cite{apple2020google}. However, we are yet to witness any major government taking their services to combat the pandemic. COVI \cite{alsdurf2020covi} is another contact tracing framework which attracted a lot of attention from academia. This work is different from the aforementioned ones in the sense that it contains discussion and mitigation strategy for ethical and privacy issues. Moreover, it also contains guidelines for secure data collection as well as using them for machine learning models which makes it the most complete and robust privacy preserving data collection and analysis framework for contact tracing data till date.

\section{PAARS Motivation and System Design}

 While designing PAARS our key motivation was to ensure maximum data security and privacy. In the meantime the utility of the system was also not to be compromised by design issues or slow adoption rate among users. Keeping these in mind, we took a novel approach in PAARS system design principles compared to already implemented systems. Instead of token exchange between users to detect proximity, we use the existing wireless network present in an organization for token exchange between users and the server which can be later used to detect contact between users. In addition to that, our system does not require any additional infrastructure or collaboration with any third party. Therefore, the scalability and adoption among users are easier than the existing approaches. 

\subsection{Detection of Contact Event between Two User}
We assume that a network consisting of wireless network access points (e.g. routers) is present in the organization environment. We denote this network as a set of access points $R=\{R_1,R_2,R_3,\cdots,R_n\}$. We assume that at any given time all elements in $R$ simultaneously broadcast a single pseudo-random number which we denote as $K_{net}$. This pseudo-random $K_{net}$ changes over time. The time difference between each new pseudo-random number broadcasted by the central network is denoted by the $\tau$.

According to Center for Disease Control guideline, a contact event between two individuals is defined as the individuals being in 2 metres of each other for more than 15 seconds \cite{cdc2020guide}. Taking this definition into account, our system divides the system environment in GPS co-ordinate blocks. A GPS co-ordinate block in our system contains co-ordinates that are within maximum 2 meters. Every co-ordinate block is assigned an unique identifier by the system. %Moreover, we assume that each GPS co-ordinate block is covered by a single wireless access point without an overlap with the neighbouring GPS co-ordinate blocks. 
\begin{figure}[!h]
    \centering
    \includegraphics[width=6cm]{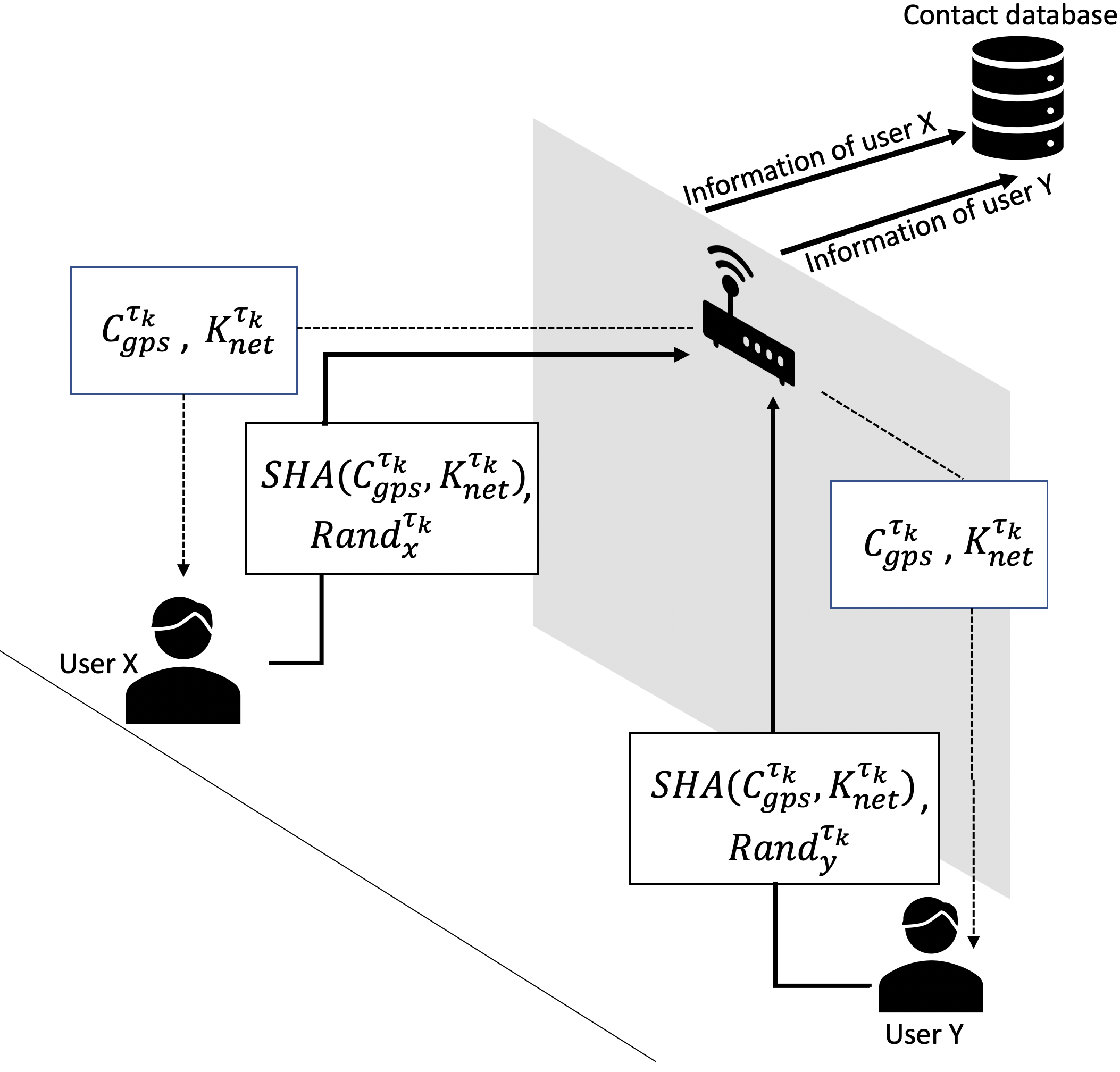}
    \caption{$User_{x}$ and $User_{y}$ is in the same GPS co-ordinate block. They connect with the same wireless access point which provides them $C^{\tau_{k}}_{gps}$ and  $K^{\tau_{k}}_{net}$. $User_x$ generates a random number $Rand_{x}^{\tau_{k}}$ and $User_{y}$ generates a random number $Rand_{y}^{\tau_{k}}$. Both users concatenate the parameters sent by the server with the network address of the access point and pass them to a hash function to generate a hashed token. Both users upload their generated random numbers and hashed tokens to the server. The server stores the hashed tokens and the random values sent by the users in a database to later use them for contact detection.}
    \label{fig:sys_design}
\end{figure}
\squeezeup
When an user enters into the system environment, the smartphone of the user connects to the network of the environment and starts exchanging tokens with the system. In order to generate a token and exchange it with the system, the user performs the following computation on the smartphone. Let's assume that the current time interval is $\tau_{k}$. Firstly, it takes the unique identifier of the GPS co-ordinate block it is currently located in. We call it $C^{\tau_{k}}_{gps}$. Secondly, it takes the pseudo-random number generated by the wireless access point covering that GPS co-ordinate block at that time which we denote as $K^{\tau_{k}}_{net}$. %Finally, it takes the network address which in this case is the IP address of the wireless access point. We call it $IP_{R_{\tau_{k}}}$. 
The token $ID$ generation process consists of concatenating these two and then passing them through SHA-256 hash function. 
\begin{center}
    $\mathrm{ID}\gets \mathrm{SHA}_\mathrm{256}(C^{\tau_{k}}_{\mathrm{gps}}||K^{\tau_{k}}_{\mathrm{net}})$
\end{center}
Therefore, it can be ensured that at a given time $\tau_{k}$, if two users are in the same block of GPS co-ordinates, they will generate the same token which will indicate a contact event between them. The user also generates a pseudo-random number $Rand_{\mathrm{user}}^{\tau_{k}}$. The user then sends this pseudo-random number along with the hashed token to the server. The server stores the hashed token, pseudo-random number , time , current infection status of this user and the probability value of being infected (initialized with value 0) in a database table. The server does not store any other persistent information about the user that might compromise the anonymity of the user. The user also stores the hashed tokens and the pseudo-random numbers sent to the servers on his/her device. 

In the server database, a contact event between two users will be registered as having same hash token values with different $Rand_{\mathrm{user}}^{\tau_{k}}$ value. 
\begin{table}[t]
    \caption{A sample contact event in the contact database. Both entries have same hashed token value but different user generated pseudo-random number entry.}
    \centering
    \begin{tabular}{|c|c|c|c|c|}
    \hline
    Hashed Token & $Rand_{\mathrm{user}}$ & Time & Status & Probability \\
    \hline
    FF56182345FA6BF7 & 18728712789 & 6:32 & N/A & 0 \\
    \hline
    FF56182345FA6BF7 & 74826390017 & 6:32 & N/A & 0 \\
    \hline
    \end{tabular}
    \label{tab:initial_entry}
\end{table}

\begin{figure}[h]
    \centering
    \includegraphics[width=\linewidth]{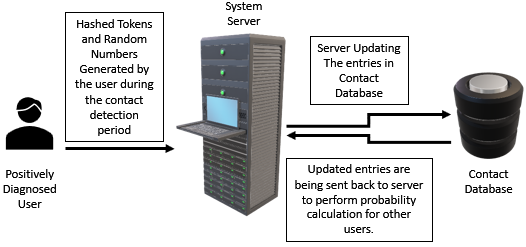}
    \caption{Workflow of the system once an user reports to be positively diagnosed and chooses to share the hashed tokens and pseudo-random numbers generated during the contact detection period of the system.}
    \label{fig:positive_diagnosis}
\end{figure}

Once an user tests positive for the infectious disease, that user can \textbf{choose to share} the hashed tokens and the random tokens generated during the contact detection phase with the system server. The integrity of the claims of positive test diagnosis can be ensured by secure methods. For instance, public health authority may provide each positively diagnosed user with an unique identifier or secure login methods to verify the claim of the user being positively diagnosed. After the authenticity of the claim of being positively diagnosed is confirmed, the user can voluntarily share the hashed tokens and the random numbers generated by the user with the system server. The system server performs a series of operations after receiving the tokens and random numbers from the user. The system updates the matching entries for the contact database where it sets the value of Status as ``Infected" and value of probability as 1. Moreover, for the matching entries of same hash tokens of other users we set the status column and probability value as ``TBD". This workflow is shown in figure \ref{fig:positive_diagnosis}. 

Referring back to table \ref{tab:initial_entry}, let's assume that an user whose claim of being positively diagnosed has been validated by the system, chose to share his/her hashed tokens and pseudo-random numbers generated with the system server. The server received all these and updated the database accordingly. Moreover, let's assume that among the contact event entries uploaded by the user, there is a contact event entry with hashed token value of ``FF56182345FA6BF7" and $\mathrm{Rand}_{\mathrm{user}}$ value of ``74826390017". After the database is updated, the relevant contact entries will be as in table \ref{tab:updated_entry}.

\begin{table}[t]
    \caption{Updated contact database entries when an user reports being infected and chooses to share relevant information with the server.}
    \centering
    \begin{tabular}{|c|c|c|c|c|}
    \hline
    Hashed Token & $Rand_{\mathrm{user}}$ & Time & Status & Probability \\
    \hline
    FF56182345FA6BF7 & 18728712789 & 6:32 & TBD & TBD \\
    \hline
    FF56182345FA6BF7 & 74826390017 & 6:32 & Infected & 1 \\
    \hline
    \end{tabular}
    \label{tab:updated_entry}
\end{table}

After updating the database entries, the server retrieves the updated entries and calculates probability of being positively diagnosed for the users who have had a contact event with the infected user. To calculate this probability we formulated a novel probability calculation model which depends on state of the art epidemiological models available in the literature such as variations of Susceptible-Infected-Removed (SIR) model  \cite{giordano2020modelling} while taking in account factors such as temporal shedding nature of the viral transmission \cite{he2020temporal}. The details of this probability calculation model is described in the following subsection.

\subsection{Mathematical Model for Probability of an User Being Positively Diagnosed}
\label{sec:probability}
Let's assume the event of being tested positive at time t is denoted by $\lambda(t)$ which is a binary random variable which takes the value 0 if the user is tested negative and 1 if tested positive. We denote the contact database entries for an user $u$ up to time $t$ as $C^{t}_{u}$. We intend to find out $\mathbf{P}(\lambda(t) = 1 | C^{t}_{u})$ which is the probability of being positively diagnosed of an user for the disease given the contact database entries of the user. Let's assume that $C^{t}_{u}$ is a set of contact entries which contains contact event entries that coincides with both infected and non-infected users. Therefore, this can be written as a set, $C^{t}_{u} =\{C^{+}_{v} + C^{-}_{v} \}$ where $v \in V$. $V$ denotes the set of users who had contact events with user $u$ and the $+$ and $-$ superscript denotes whether the user $v$ is diagnosed as being positively infected or not. The set of all positively diagnosed users are denoted as $D^+$. Therefore, we can write the intended probability function of being infected as: 
\begin{equation}
    \mathbf{P}(\lambda(t) = 1 | C^{t}_{u}) = \mathbf{f}(\mathbf{P}(\lambda(t) = 1 | C^{+}_{v}));  \forall v \in (V \cap D^+)
    \label{eq:eq1}
\end{equation}

To calculate our intended probability function $\mathbf{f}$, we need to find the average of the probability of each individual contact event with user $v$ who were positively diagnosed for the disease. This is due to the fact that each contact event with a positively diagnosed user $v \in V$ can contribute to the user $u$ in question to be positively diagnosed. Moreover, we assume that each contact event between two users are independent of any prior contact events between any two users. Therefore, it is possible to view each individual contact events between users as independent events. Therefore, the equation \ref{eq:eq1} can be written as:  
\begin{equation}
\begin{split}
    \mathbf{P}(\lambda(t) = 1 | C^{t}_{u}) = \frac{1}{N} \sum_{i=0}^N \mathbf{P}(\lambda(t) = 1 | C^{+}_{v});\forall v \in (V \cap D^+) \\
    \qquad where, N = |(V \cap D^+)|
\end{split}
    \label{eq:eq2}
\end{equation}

In order to understand how our system calculates the individual probability of each contact event contributing to the total probability we would like to draw attention to the \cite{giordano2020modelling} which shows how a variant of widely used Susceptible-Infectious-Removed (SIR) model was used to mathematically model the recent COVID-19 pandemic outbreak in Italy. In this work, the authors noted that transmission of the disease from one individual to another can depend on multiple factors. However, the most important factors are the time duration of a contact event between two individuals and the viral shedding factor of the infected individual. The later one can be described as the rate at which the infected individual is spreading the virus within a certain distance \cite{cdc2020guide}. We take these two important aspect into account and design our probability calculation method accordingly. 

We define $E^v_u(t)$ as the exposure score of the contact event between user $u$ and $v$ at time $t$ which we calculate from the work described in \cite{sattler2020risk}. In this work, the authors design multiple models to determine the severity of a contact event between two individuals. In our particular case, we will use the sigmoid model developed by the authors to determine the severity of a contact event. This model takes the duration of a contact and the distance between two individuals during the contact as input of the model and produces a scalar value which is the severity score of the contact event. 

We define $S^v(t)$ as the viral shedding rate of the infected user at time t. This can be directly determined by using the work described in \cite{he2020temporal}. In this work, the authors drew the temporal relation between the days of infection and viral shedding. According to this work, for the first few days after being infected, the viral shedding of an infected person is significantly higher than the subsequent days. We can get $S^v(t)$ from this model by inputting the number of days when the contact event took place.

Therefore, we can derive the probability of being positively diagnosed from each contact event by using the following formulation:
\begin{equation}
    \mathbf{P}(\lambda(t) = 1 | C^{+}_{v}) = \frac{E^v_u(t) * S^v(t)}{\max(E^v_u(t) * S^v(t))}
    \label{eq:eq3}
\end{equation}

Using equation \ref{eq:eq3}, we can calculate the probability of being positively diagnosed for each contact events and average them to get absolute probability score for an user being positively diagnosed. However, there is a caveat in the above formulation which is privacy breach. A malicious user may track the differences between the updates of his/her probability score and might attempt to infer which contact event caused the update of the probability score. Therefore, we attempt to mitigate this problem by using differential privacy. After calculation of the left hand term in \ref{eq:eq3}, we add a noise to it which is from Laplace distribution. Then equation (\ref{eq:eq3}) becomes:
\begin{equation}
    \mathbf{P}(\lambda(t) = 1 | C^{+}_{v}) = \frac{E^v_u(t) * S^v(t)}{\max(E^v_u(t) * S^v(t))} + \mathcal{L}(0, \frac{\alpha}{N})
    \label{eq:eq4}
\end{equation}
where $\alpha$ is the regularization parameter for the noise and the mechanism is $\frac{N}{\alpha}$ differentially private. The accuracy analysis of this noise addition mechanism will be discussed in the utility analysis section.

Using equation \ref{eq:eq4} for all the contact events between the user in question $u$ and the infected users $v$, we can derive noisy probability scores for each contact event. We input these noisy probability scores in equation \ref{eq:eq2} and calculate the final probability score of an user being positively diagnosed given the contact events in the contact database.

\subsection{Privacy Aware User Alerting System}
The server has a set of hashed tokens $H_{sys}$ which contains the hashed tokens that have positive probability scores. Let's assume user $u$ wants to access his/her probability score from the system. In order to do that, the user uses $H_{u}$ which is the relevant hashed tokens within the time period which is generated by the user and have been shared to the server. The server and the user performs a PSI (private set intersection) as described in the work of \cite{trieu2020epione}. The result of $PSI(H_{sys}, H_{u})$ is shared to the user $u$. If the result returns a non-empty set, the user $u$ makes a request to the server which contains the result of the $PSI(H_{sys}, H_{u})$ and the $Rand_{user}^{\tau_{k}}$ associated with each elements in the result of $PSI(H_{sys}, H_{u})$. The server receives the request and sends back the probability scores associated with the query. The user receives the result and based on the result the user can decide to self-isolate, quarantine or seek medical help from professionals. It is important to note that the user gets the probability scores associated with the entries in the result of $PSI(H_{sys}, H_{u})$ and  his/her generated $Rand_{user}^{\tau_{k}}$ only. Therefore, he/she does not have access to the probability scores calculated for the other users with the same hashed token values.

\section{Utility Analysis}
\subsection{Building Occupancy Level Determination:} PAARS can automatically calculate the building occupancy level in real time by analyzing the number of active connections sending in hashed tokens in every wireless access point within the system environment. Based on the level of occupancy at a given time, it can alert the authorities if maximum occupancy level set by health professionals are being violated.
\subsection{Error Estimation of the Probability Calculation Model:} In order to calculate the accuracy of the noisy probability scores in subsection \ref{sec:probability}, we adopt the accuracy analysis method presented in the work of \cite{cao2013efficient} and \cite{hay2009boosting}. Let's assume $P_{actual}$ is the actual probability and $P_{noisy}$ is the calculated noisy probability. According to \cite{hay2009boosting}, the error of a differentially private mechanism output is determined by the variance $V(P) = \mathop{\mathbb{E}}[(P_{noisy}-P_{actual})^2 ]- \mathop{\mathbb{E}}[(P_{noisy}-P_{actual}) ]$, where $\mathop{\mathbb{E}}$ is the mathematical expectation. Referring to equation \ref{eq:eq4}, $(P_{noisy}-P_{actual})$ in this case is $\mathcal{L}(0, \frac{\alpha}{N})$ which has variance of $\frac{2\alpha^2}{N^2}$. Therefore, each term on the right hand side in equation \ref{eq:eq2} is expected to add $\frac{2\alpha^2}{N^2}$ error to the final output. Accordingly, the final expected accumulated error is:
\begin{equation}
    error = \frac{1}{N} \sum_{i=0}^N \frac{2\alpha^2}{N^2} = \frac{2\alpha^2}{N^2}
    \label{eq:eq5}
\end{equation}

\section{Security Analysis}
\label{sec:sec_analysis}
 To analyze the security of the system firstly we consider a set of parties and a few protocols. We assume that in our proposed system there is a set of parties who have agreed to perform some computation. Moreover, the parties have consented to release the final result of the computation to a specific party and nothing else will be released from any other parties, not even the computational process used. There are two classical security models in this case: 

\begin{itemize}
    \item Semi-Honest Model: This type of adversary is someone who is presumed to follow the execution protocol. However, it attempts to obtain extra information from the execution protocol.
    \item Malicious Model: This type of adversary may attack the system using any possible strategy i.e. brute force attack, side channel attack, supplying inconsistent output or trying to infer information about system data or execution process of the computation involved.
\end{itemize}
In this work, we define three individual entities: The user group $U$, the semi-honest system server $S$ and malicious adversary $A$ who would try to attack the system using any possible method that is executable in polynomial time. Attacker $A$ can be a malicious user or someone outside the system environment. For simplicity, we assume that the no entity is colluding with one another. Moreover, we assume that all communications between the user $U$ and the server $S$ are securely authenticated (i.e. TLS).

The robustness of all proposed decentralized solutions so far relies on the fact that individual user identities are not linkable to the pseudo-random tokens they generate. However, the exchange of these pseudo random tokens between users still poses some security threats such as possibility of identifying positively tested users in case of a single contact event in the given time period. In the following paragraphs we will show that our proposed system prevents majority of the security risks of the previous approaches while ensuring privacy.
\begin{itemize}
    \item \textbf{Positively tested user identification attack}: In this attack a malicious user tries to identify a positively tested user by matching the tokens exchanged between them. Most of the proposed contact tracing/proximity monitoring systems have token exchange as a core part of their functionality. However, in our proposed system there is no token exchange between two users. Therefore, this attack is not feasible in our system.
    \item \textbf{False positive reporting attack by user}: In this attack a malicious user $U_{mal}$ tries to generate false alarm, by reporting him/her self as positively diagnosed with the disease to the server $S$. This attack could cause erroneous computation on the server and result in confusion and panic among users. However, in our proposed system, this attack is not possible as every claim made by any user of being positively diagnosed for the disease is verified with the proper authority.
    \item \textbf{Impersonation attack by user:} This attack is aimed at faking a person's presence when that person was actually not present there. The attacker $A$ gathers exchanged tokens from other users and broadcasts them. If any of those tokens are later marked as tokens from an individual who has been positively diagnosed with the disease, a number of users may be falsely alerted of being in contact with that person while in reality they never had any contact with that positively diagnosed person. In our proposed system this attack is not possible as individual users do not exchange tokens with each other.
    \item \textbf{Exposure of Social Interaction Graph of the user by the server:} One of the major privacy issues for any contact tracing or access monitoring/regulations system is that there are inherent risks of the server $S$ associating an user $U$ with a set of users $U_{contacts}$ where each member of the set $U_{contacts}$ is the ones who were in proximity of $U$. In our proposed system, the server $S$ does not keep any persistent identifier of any user $U$ and the user $U$ never shares any persistent identifier of him/her to the server $S$. The server $S$ only receives the securely hashed values from the user $U$ which can not be linked back to individual users. Therefore, in our proposed system this attack is not possible.
    
    \item \textbf{Track user's location by the server:} This attack is performed by the server to use the shared tokens to track the location of any individual user. Theoretically speaking, any system where the user $U$ exchanges some kind of token with the server $S$ , it is possible for the server $S$ to track the user by using passive packet sniffer. In our proposed system we assumed the server to be semi-honest entity which does not employ any such malicious tactics. Therefore, it is not possible for the server $S$ in our proposed system to track the users movement since the exchanged tokens contain no persistent identifier and they change over time based on user's location.
\end{itemize}

\section{Discussion}
\subsection{Decentralized Data Storage:} One of the key differences between PAARS and the existing approaches is the fact that PAARS enables decentralized data storage. Each organization implements their own version of PAARS and stores the data locally. Therefore, it is not feasible for an adversary to compromise all different data storages residing in different organizations at once and get a complete picture. On the contrary, majority of the existing approaches that rely on ephemeral token exchange between users store these tokens from all users in large central databases. Therefore, a data breach in these systems from the inside or the outside can bear catastrophic effect in terms of privacy. 

\subsection{Differential Privacy:} To the best of our knowledge, no work in literature yet addressed the user side privacy issues in contact tracing/access monitoring systems by incorporating differential privacy. Most of these work rely on using homomorphic encryption or private set intersection (PSI) method \cite{trieu2020epione} which is computationally costly. In our work, we use efficient PSI for once in the user query only. The rest of the privacy is guaranteed by differential privacy.

\subsection{Infrastructure Requirement:} \label{sec:infrastructure} PAARS do not require any significant addition in the infrastructure available in the environment. This is a major advantage over existing systems or the ones being implemented. For instance, every organization has wireless network now a days and majority of the population has access to smartphones. Therefore, deploying and maintaining PAARS is easy and adoption should be fast. On the contrary, implementing PAARS architecture in city-wide scale would require a singular network which has city-wide coverage. Cell service providers are capable of providing such network. However, to collaborate with them to build and deploy a system like this would require a lot of time and effort, whereas, PAARS can be deployed quickly in organizational level. 
\subsection{GPS accuracy:} GPS accuracy can be perceived as a technical issue in PAARS system design. However, using other sensors (i.e. magnetometer and accelerometer and IMU) available in smartphones, the accuracy issue can be effectively resolved \cite{poulose2019indoor} . 

\section{Conclusion}
In this paper, our main contribution is to show that the privacy issues of contact tracing/access monitoring systems can be efficiently solved if the problem is handled in a decentralized and micro scale. By designing PAARS, we have shown that our approach can be used as a framework for reducing viral transmission risks in organizations while simultaneously allowing them to return to normal operational capacity. As a future work, we plan to implement the system and analyze it's performance and other relevant issues in terms of privacy and efficiency in system design. Another direction of farther extension of the work described in this paper, is the analysis of the novel probability calculation method we proposed. By implementing rapidly changing disease models available in the literature, we hope to improve the performance of our probability calculation method.
\section*{Acknowledgments}
This research was supported in part by the NSERC Discovery Grants (RGPIN-2015-04147). 
\squeezeup


%\section*{Acknowledgment}




%\bibliographystyle{IEEEtranS}
\bibliographystyle{plain}
\bibliography{Bibliography}

\begin{thebibliography}{10}

\bibitem{cdc2020guide}
{\em CDC guideline}, 2020 (accessed Aug 22, 2020).
\newblock
  https://www.cdc.gov/coronavirus/2019-ncov/prevent-getting-sick/social-distancing.html.

\bibitem{korea2020ct}
{\em Seoul's Radical Experiment in digital contact tracing}, 2020 (accessed
  August 21, 2020).
\newblock
  https://www.newyorker.com/news/news-desk/seouls-radical-experiment-in-digital-contact-tracing.

\bibitem{apple2020google}
{\em Apple and Google releases contact tracing API's for building mobile apps},
  2020 (accessed June 3, 2020).
\newblock
  https://www.apple.com/newsroom/2020/04/apple-and-google-partner-on-covid-19-contact-tracing-technology.

\bibitem{dp20203t}
{\em DP-3T}, 2020 (accessed June 3, 2020).
\newblock https://github.com/DP-3T/documents.

\bibitem{pepp20203t}
{\em PEPP-3T}, 2020 (accessed June 3, 2020).
\newblock https://www.pepp-pt.org.

\bibitem{alsdurf2020covi}
Hannah Alsdurf, Yoshua Bengio, Tristan Deleu, Prateek Gupta, Daphne Ippolito,
  Richard Janda, Max Jarvie, Tyler Kolody, Sekoul Krastev, Tegan Maharaj,
  et~al.
\newblock Covi white paper.
\newblock {\em arXiv preprint arXiv:2005.08502}, 2020.

\bibitem{altuwaiyan2018epic}
Thamer Altuwaiyan, Mohammad Hadian, and Xiaohui Liang.
\newblock Epic: efficient privacy-preserving contact tracing for infection
  detection.
\newblock In {\em 2018 IEEE International Conference on Communications (ICC)},
  pages 1--6. IEEE, 2018.

\bibitem{bay2020bluetrace}
Jason Bay, Joel Kek, Alvin Tan, Chai~Sheng Hau, Lai Yongquan, Janice Tan, and
  Tang~Anh Quy.
\newblock Bluetrace: A privacy-preserving protocol for community-driven contact
  tracing across borders.
\newblock {\em Government Technology Agency-Singapore, Tech. Rep}, 2020.

\bibitem{berke2020assessing}
Alex Berke, Michiel Bakker, Praneeth Vepakomma, Ramesh Raskar, Kent Larson, and
  AlexSandy' Pentland.
\newblock Assessing disease exposure risk with location histories and
  protecting privacy: A cryptographic approach in response to a global
  pandemic.
\newblock {\em arXiv preprint arXiv:2003.14412}, 2020.

\bibitem{cao2013efficient}
Jianneng Cao, Qian Xiao, Gabriel Ghinita, Ninghui Li, Elisa Bertino, and
  Kian-Lee Tan.
\newblock Efficient and accurate strategies for differentially-private sliding
  window queries.
\newblock In {\em Proceedings of the 16th International Conference on Extending
  Database Technology}, pages 191--202, 2013.

\bibitem{chan2020pact}
Justin Chan, Shyam Gollakota, Eric Horvitz, Joseph Jaeger, Sham Kakade,
  Tadayoshi Kohno, John Langford, Jonathan Larson, Sudheesh Singanamalla, Jacob
  Sunshine, et~al.
\newblock Pact: Privacy sensitive protocols and mechanisms for mobile contact
  tracing.
\newblock {\em arXiv preprint arXiv:2004.03544}, 2020.

\bibitem{dar2020applicability}
Aaqib~Bashir Dar, Auqib~Hamid Lone, Saniya Zahoor, Afshan~Amin Khan, and
  Roohie~Naaz Mir.
\newblock Applicability of mobile contact tracing in fighting pandemic
  (covid-19): Issues, challenges and solutions.
\newblock {\em IACR Cryptol. ePrint Arch.}, 2020:484, 2020.

\bibitem{giordano2020modelling}
Giulia Giordano, Franco Blanchini, Raffaele Bruno, Patrizio Colaneri,
  Alessandro Di~Filippo, Angela Di~Matteo, and Marta Colaneri.
\newblock Modelling the covid-19 epidemic and implementation of population-wide
  interventions in italy.
\newblock {\em Nature Medicine}, pages 1--6, 2020.

\bibitem{hay2009boosting}
Michael Hay, Vibhor Rastogi, Gerome Miklau, and Dan Suciu.
\newblock Boosting the accuracy of differentially-private histograms through
  consistency.
\newblock {\em arXiv preprint arXiv:0904.0942}, 2009.

\bibitem{he2020temporal}
Xi~He, Eric~HY Lau, Peng Wu, Xilong Deng, Jian Wang, Xinxin Hao, Yiu~Chung Lau,
  Jessica~Y Wong, Yujuan Guan, Xinghua Tan, et~al.
\newblock Temporal dynamics in viral shedding and transmissibility of covid-19.
\newblock {\em Nature medicine}, 26(5):672--675, 2020.

\bibitem{poulose2019indoor}
Alwin Poulose, Odongo~Steven Eyobu, and Dong~Seog Han.
\newblock An indoor position-estimation algorithm using smartphone imu sensor
  data.
\newblock {\em IEEE Access}, 7:11165--11177, 2019.

\bibitem{raskar2020apps}
Ramesh Raskar, Isabel Schunemann, Rachel Barbar, Kristen Vilcans, Jim Gray,
  Praneeth Vepakomma, Suraj Kapa, Andrea Nuzzo, Rajiv Gupta, Alex Berke, et~al.
\newblock Apps gone rogue: Maintaining personal privacy in an epidemic.
\newblock {\em arXiv preprint arXiv:2003.08567}, 2020.

\bibitem{sattler2020risk}
Felix Sattler, Jackie Ma, Patrick Wagner, David Neumann, Markus Wenzel, Ralf
  Sch{\"a}fer, Wojciech Samek, Klaus-Robert M{\"u}ller, and Thomas Wiegand.
\newblock Risk estimation of sars-cov-2 transmission from bluetooth low energy
  measurements.
\newblock {\em arXiv preprint arXiv:2004.11841}, 2020.

\bibitem{trieu2020epione}
Ni~Trieu, Kareem Shehata, Prateek Saxena, Reza Shokri, and Dawn Song.
\newblock Epione: Lightweight contact tracing with strong privacy.
\newblock {\em arXiv preprint arXiv:2004.13293}, 2020.

\end{thebibliography}

\end{document}